\documentclass[conference,9pt]{IEEEtran}
\IEEEoverridecommandlockouts

\usepackage{cite}
\usepackage{amsmath,amssymb,amsfonts}
\usepackage{algorithmic} 
\usepackage{graphicx}
\usepackage{multirow}
\usepackage{textcomp}
\usepackage{xcolor}
\usepackage{booktabs}
\usepackage{hyperref}

\def\BibTeX{{\rm B\kern-.05em{\sc i\kern-.025em b}\kern-.08em
    T\kern-.1667em\lower.7ex\hbox{E}\kern-.125emX}}
\begin{document}

\title{Unified Pathological Speech Analysis with Prompt Tuning
}

\author{
\IEEEauthorblockN{Fei Yang, Xuenan Xu,  Mengyue Wu$\dag$, Kai Yu$\dag$ \thanks{$\dag$Mengyue Wu and Kai Yu are the corresponding authors.}}
\IEEEauthorblockA{
    \textit{MoE Key Lab of Artificial Intelligence, AI Institute}\\
    \textit{
     X-LANCE Lab, Department of Computer Science and Engineering} \\
     \textit{Shanghai Jiao Tong University
     }\\
Shanghai, China \\
\{fyang20528, wsntxxn, mengyuewu, kai.yu\}@sjtu.edu.cn}

}
\maketitle

\begin{abstract}
Pathological speech analysis has been of interest in the detection of certain diseases like depression and Alzheimer's disease and attracts much interest from researchers. However, previous pathological speech analysis models are commonly designed for a specific disease while overlooking the connection between diseases, which may constrain performance and lower training efficiency.
Instead of fine-tuning deep models for different tasks, prompt tuning is a much more efficient training paradigm. We thus propose a unified pathological speech analysis system for as many as three diseases with prompt tuning technique. This system uses prompt tuning to adjust only a small part of the parameters to detect different diseases from speeches of possible patients. Our system leverages a pre-trained spoken language model and demonstrates strong performance across multiple disorders while only fine-tuning a fraction of the parameters. This efficient training approach leads to faster convergence and improved F1 scores by allowing knowledge to be shared across tasks. Our experiments on Alzheimer’s disease, Depression, and Parkinson's disease show competitive results, highlighting the effectiveness of our method in pathological speech analysis. 
\end{abstract}


\section{Introduction}

A speech signal contains a wealth of information about the speaker, including the linguistic content of their message and paralinguistic aspects like their emotion, age, and gender~\cite{CUMMINS201841}.
Analyzing speech signals has proven highly effective in diagnosing conditions such as Alzheimer's disease, depression, and Parkinson's disease.
By examining various acoustic and prosodic features of a patient's voice, such as speech rate, pitch, intensity, and fluency, it is possible to detect and differentiate these illnesses~\cite{koops2023speech}. Many studies focus on extracting various acoustic features from participants' speech and applying machine learning techniques for the detection of different conditions~\cite{ZOLNOORI2023102624,liu2023efficient,liu2023ensemble,lamba2023speech,jahan2024early}.
These studies identify discriminative features within speeches and have demonstrated promising performance in detection tasks.

With the success of deep learning, there has been a proliferation of works utilizing neural networks to construct such detection systems, and they have shown promising results.
For example, Bertini \textit{et al.}~\cite{bertini2022automatic} proposed an autoencoder-based method for Alzheimer's disease classification from speech, using a recurrent neural network (RNN) based model for unsupervised feature extraction.
Rejaibi \textit{et al.}~\cite{REJAIBI2022103107} proposed a deep RNN  based framework for depression detection from speech and demonstrated good results.
Yin \textit{et al.}~\cite{yin2023depression} proposed a deep learning model that combines a parallel convolutional neural network (parallel-CNN) and Transformer to effectively capture local and temporal sequential information with manageable complexity.
Quan \textit{et al.}~\cite{quan2022end} proposed an end-to-end deep learning model for Parkinson's disease detection from speech, combining 2D-CNNs for dynamic feature extraction and 1D-CNNs for temporal dependencies.
The application of deep learning techniques to pathological speech analysis has greatly enhanced the accuracy and reliability of disease detection systems. These improvements underscore the potential of neural networks in this field and set the stage for continued research and development in speech-based disease diagnosis.

\begin{figure}[t]
  \centering
  \includegraphics[width=\linewidth]{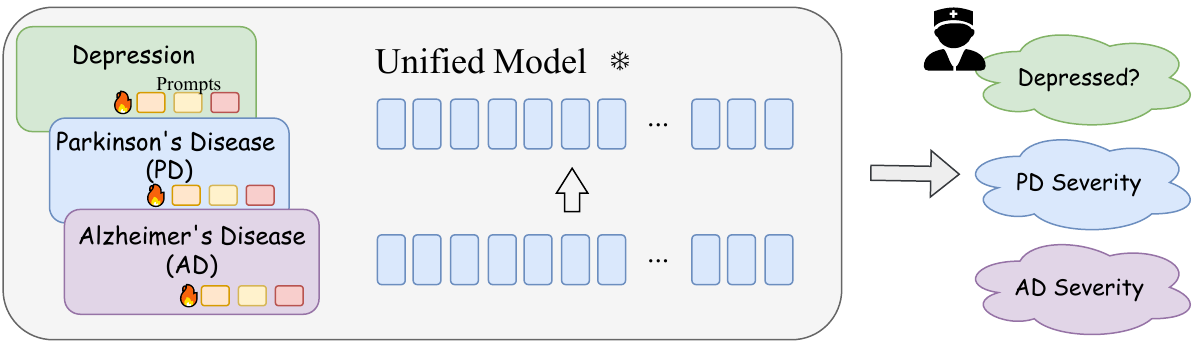}
  \caption{The overview of our proposed system. We use prompts to guide the unified model to detect different diseases.}
  \label{fig:intro}
\end{figure}

However, current models are typically trained on datasets focused on specific diseases, which limits their ability to generalize to other conditions.
This lack of flexibility makes it challenging to directly apply these models to different disease detection tasks. 
To overcome this limitation, we need approaches that can leverage shared knowledge across various disease types, allowing models to handle multiple conditions more effectively.
A traditional approach is fine-tuning, where a model trained on one disease dataset is further fine-tuned on data from a different disease. This process requires human experts to decide which part of the model's parameters should be adjusted, often involving a careful selection of layers to freeze or modify. The learning rate schedule, along with the amount of new data, are also key factors. However, these settings tend to vary greatly across tasks, making fine-tuning less suitable for unified pathological speech analysis, as it struggles to generalize across multiple diseases without disease-specific adjustments.

Prompt tuning is an emerging technique used to enhance the performance of pre-trained models. In the context of language models (LMs), a prompt refers to the initial input or context provided to the model, which guides its understanding and generation~\cite{10.1145/3560815}. By updating a limited number of continuous prompt vectors, prompt tuning offers a more efficient way to adapt pre-trained models to specific tasks.
One popular approach within this framework is prefix-tuning~\cite{li-liang-2021-prefix}, which focuses on deep prompt tuning by adding trainable prompts to the input of the LM. This method leverages the shared knowledge across various tasks embedded within the pre-trained model, making it particularly useful in cases where labeled data is scarce. Prefix-tuning not only avoids the need to train separate models for each new task or dataset but also improves generalization across tasks, enhancing training efficiency.

In recent years, prompt tuning has been applied to speech processing tasks, demonstrating promising results. For example, SpeechPrompt~\cite{chang2022speechprompt} and SpeechPrompt v2~\cite{chang2023speechprompt} use the Generative Spoken Language Model (GSLM)~\cite{lakhotia2021generative} as their backbone and employ prompt tuning to create a unified speech analysis framework. These approaches have shown that prompt tuning can deliver strong performance on small speech datasets across multiple analysis tasks. Building on these advancements, we adapt prompt tuning specifically for pathological speech analysis. By applying this method, we aim to explore its potential for improving diagnostic accuracy and efficiency in specialized speech disorder scenarios, further highlighting the versatility and power of prompt-tuned models in speech processing.

To our knowledge, this is the first time that prompt tuning is applied to pathological speech analysis.
The main contributions of this work are:
\begin{itemize}
\item Propose a pathological speech analysis framework that can unify different \textbf{\textit{diseases, languages and class labels}}, which is the first piece to use prompt tuning in this task.
\item Utilize pre-trained LMs to do pathological audio detection, encoding audios into tokens and using prompt tuning to make LMs capable of detecting different kinds of diseases.
\item Perform experiments on three diseases, namely Alzheimer's disease (AD), Parkinson's disease (PD) and Depression encompassing three different datasets, results indicating that our model performs well across all disease tasks, demonstrating its robustness in handling diverse pathological speech datasets.
\end{itemize}


\section{Methodology}
In this section, we describe our proposed framework for using prompt tuning in pathological speech analysis.
Our framework is built upon GSLM~\cite{lakhotia2021generative} and consists of three key components.
Firstly, a speech-to-unit module converts the input speech signal into units that can be processed by the LM.
Secondly, we employ a prompt tuning approach with a fine-tuned uLM (unit Language Model).
By leveraging a pre-trained uLM, we adapt it through prompt tuning to effectively handle different disease datasets.
This enables the model to generate contextually appropriate responses for various medical conditions.
Lastly, we incorporate a verbalizer to establish mappings between the tokens generated by the uLM and classification labels.
This facilitates the interpretation and output alignment of the model with the specific disease classification results.
\begin{figure}[t]
  \centering
  \includegraphics[width=\linewidth]{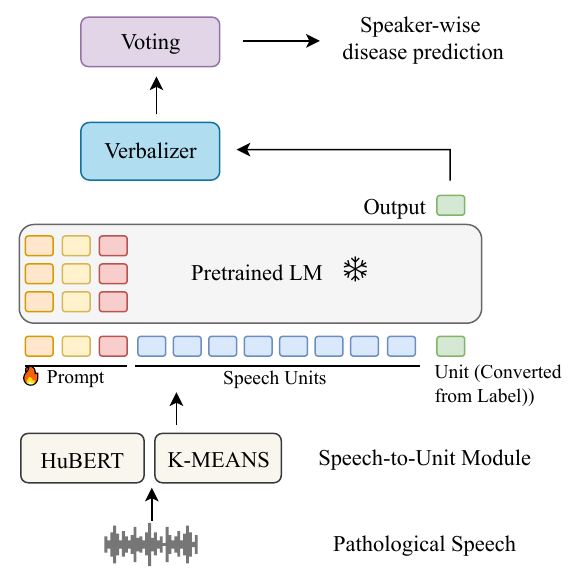}
  \caption{Schematic diagram of the unified pathological speech analysis framework. We take speech from datasets covering different diseases, languages, class labels and leverage these metadata as trainable prompts.}
  \label{fig:framework}
\end{figure}

\subsection{Speech-to-Unit Module}
Speech self-supervised learning is effective for feature extraction in speech because it leverages the abundance of unlabeled data to capture both acoustic and linguistic information, enabling the model to learn meaningful representations that capture the inherent structure and properties of speech.

HuBERT~\cite{hsu2021hubert} is a speech representation model.
It combines the strengths of both supervised pre-training and self-supervised learning to create a powerful and versatile model for speech processing tasks.
HuBERT is trained using a self-supervised learning approach on a large amount of multilingual and multitask speech data, including both labeled and unlabeled data. The model learns useful speech representations without requiring labeled data during pre-training, which makes it effective for a wide range of downstream tasks.
As a result, we choose to use HuBERT to generate units for the LM.

We use k-means clustering to convert the continuous frame representations learned by HuBERT into discrete representations, which are then used as input units for the uLM.

\subsection{Unit-Language Model with Prompt Tuning}

We use the pre-trained uLM from GSLM, which excels at modeling the rich expressiveness of spoken language based on raw audio units without relying on text or labels. This model is based on the Transformer architecture, consisting of 12 layers, 16 attention heads, an embedding size of 1024, and a feed-forward network size of 4096, with a dropout probability of 0.1. 
In our prompt-tuning approach, we construct prompts based on specific disease and dataset information. We employ two strategies: First, we use the prompt as a prefix to the input sequence for the uLM, which includes the prompt and the unit sequence generated by the speech-to-unit module. Second, we apply deep prompt tuning by adding the prompts as prefixes within all Transformer blocks of the uLM. 

During the training process, we employ a strategy where we freeze the majority of the model's parameters, except for those specifically related to the prompt.
This allows us to isolate and focus the model's learning on adjusting the prompt-related parameters.
By doing so, we fine-tune the model's behavior and responses to align with the desired output criteria, while leveraging the pre-trained knowledge and capabilities of the uLM.


The prompt is optimized by minimizing the language modelling loss:
\[
\mathcal{L}_{\text{task}}(S; P, \theta) = - \frac{1}{N} \sum_{i=1}^{N} \sum_{c=1}^{C} y_{i}^{c} \log \hat{y}_{i}^{c},
\]
where $P$ is the prompt sequence, $S$ is the speech unit sequence, and $\theta$ denotes the frozen uLM parameters.
$N$ is the data size.
$y_{i}^c$ and $\hat{y}_{i}^c$ represent the ground truth label and predicted probability of the $c$-th class for sample $i$.


In our approach to prompt tuning, we employ a carefully structured prefix prompt designed to guide the inference process of the unit language model. 
This prefix prompt is made up of three key components: disease-specific embeddings, language embeddings, and class embeddings.
These components work together to provide the model with a rich contextual foundation, enabling it to generate outputs that are not only relevant but also aligned with the specific requirements of the task.

\begin{itemize}
\item \textbf{Disease-Specific Embedding Prompts}: These embeddings are designed to capture the nuances of the specific detection task for which the model is being fine-tuned.
By embedding disease-related information directly into the prompt, we ensure that the model is primed to process the input sequences in a way that is highly relevant to the task context from the outset.
This helps focus the model's inferential capabilities on the specific requirements and constraints of the task, thereby improving the accuracy and relevance of the generated speech tokens.

\item \textbf{Language Embedding Prompts}: Language embeddings play a crucial role in guiding the model to generate speech tokens that adhere to the phonetic patterns of the target language, for in real-world pathological speech analysis there are often speeches in different languages. 
By incorporating these embeddings into the prefix prompt, we provide the model with a strong ability to deal with speeches in different languages.
This ensures that the generated tokens not only meet the task requirements but also maintain fluency and consistency in the target language.

\item \textbf{Class Embedding Prompts}: 
In different diseases and datasets, the output class varies depending on the specific task. For instance, in one dataset, the classes may represent different mental health conditions, while in another, they might represent emotional states. Class embeddings encode information about these categories or labels relevant to the disease.
These prompts direct the model's attention to specific classes.
By including these embeddings in the prefix, we enhance the model's ability to produce contextually appropriate outputs that align with the desired class outcomes.
\end{itemize}


By structuring the prefix prompt in this manner, we leverage the pre-trained knowledge of the underlying model while embedding it with specific contextual information.
This organization allows the model to process the input in a way that integrates multiple layers of context (disease-specific, linguistic, and categorical) before generating the output.
This approach allows the model to output units that are coherent and closely aligned with the specific goals of the task by providing relevant context directly through the prompt.

\begin{figure}[t]
  \centering
  \includegraphics[width=0.95\linewidth]{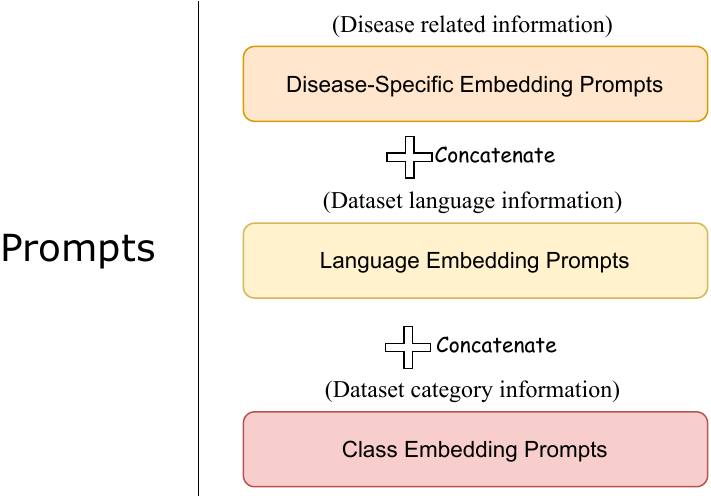}
  \caption{We use prefix prompt to direct the unit language model's inference. Three main elements are involved: disease-specific, language, and class embeddings. Together, they offer a comprehensive context, allowing the model to produce task-specific and relevant outputs.}
  \label{fig:prompt}
\end{figure}

\subsection{Verbalizer}
A verbalizer is an important component that facilitates the mapping between the output tokens generated by the language model and their corresponding classification labels.
It helps establish a connection and alignment between the model's internal representation and the desired external representation.
In most cases, the design of a verbalizer involves manual effort to establish an improved mapping relationship between the tokens generated by the LM and the task-specific labels.
To make our framework automatic and minimize the need for additional human effort, we utilize a fully connected layer to learn the verbalizer.
This approach enables the framework to automatically learn the intricate mapping relationships, thereby reducing the reliance on manual design and facilitating a higher degree of automation.

\subsection{Voting}
Since the training audio recordings are typically long and only have one overall label for the entire recording, segmenting long audio into smaller segments for training purposes is necessary.
However, this may introduce certain challenges in the training data.
Some segments labeled as positive may not exhibit strong pathological features and may even be confused with normal speech.
To address this issue, we employed a voting mechanism in which the final classification result is determined by aggregating the individual classification results of each segment through voting. We use it in the inference phase after training.
This approach allows for a more robust and consensus-based decision-making process, taking into account the collective judgment of multiple segments rather than relying solely on the characteristics of individual segments.
By leveraging the voting strategy, we aim to enhance the accuracy and reliability of the classification results in the presence of potential ambiguities or uncertainties in the segmented training data.
The voting is done in the following way:

\begin{equation}
    C_i = \begin{cases} 1, & \text{when} \sum_{j=1}^{n_i} (C^i_j ) > \rho n_i \\ 0, & \text{when} \sum_{j=1}^{n_i} (C^i_j ) \leq \rho n_i \end{cases},
\end{equation}
where ${C_i}$ stands for the classification result of the i-th patient's speech, ${C^i_j}$ stands for the j-th segment in the i-th speech, $\rho$ stands for a ratio constant and ${n_i}$ stands for the number of segments.

\section{Experiments}
In order to test the performance of our framework, we conduct experiments on three different disease datasets.

\subsection{Datasets}
The \textbf{MDD} corpus is an extensive conversational dataset specifically developed for the detection of major depressive disorder (MDD)~\cite{DI2021161}.
It includes 1000 hours of speech conversations between interviewers and subjects, involving 588 healthy subjects and 545 subjects diagnosed with depression.
To focus solely on the subject's content, we perform speaker diarization, ensuring that only the subject's utterances are retained.
These individual utterances are resampled to 16kHz, then concatenated and segmented into 5-second intervals.
This dataset offers valuable insights for the identification and study of MDD.

The \textbf{AD} corpus is a Mandarin speech dataset consisted of about 400 hours of interview recordings between elderly participants and doctors collected by Shanghai Mental Clinic Center~\cite{li2019building}.
Like the pre-process steps for the MDD corpus, we also apply speaker diarization to extract speeches from participants.
We concatenate these utterances and cut them into segments of fixed 5-second length.
We treat every segment as an individual sample and give them labels as the original file.
This dataset is a big dataset and have great value in Mandarin AD speech detection.

The \textbf{PD} dataset is a Mandarin speech dataset consisting of speeches from 380 patients with Parkinson's disease.
The total length of audios is about 12.6 hours.
In this dataset, there are audio files of patients reading weather forecasts under two different conditions: with medicine taken and without medicine taken.
In our experiments, we label the drug-taken samples as negative samples, and the rest are labeled as positive samples.
The audio files are cut into 5-second length segments for model training.

\subsection{Experimental Setup}

The experiments were implemented under the Fairseq framework~\cite{ott2019fairseq}. All experiments were conducted on a single A10 GPU. In our experiments, training was conducted using the Adam optimizer with a learning rate of $5\times10^{-2}$.
The learning rate was scheduled to decrease by a factor of 0.1 when the validation loss plateaued for 5 consecutive epochs. We trained the model for a maximum of 300 epochs, with early stopping applied if the performance failed to improve for 15 epochs. To accelerate training, we used FP16 mixed precision. 

\subsection{Results}

\begin{table}[th]
  \caption{Experimental Results and Baseline Comparison}
  \label{tab:result}
  \centering
  \begin{tabular}{ c c c c }
    \toprule
    \textbf{Dataset} & \textbf{Metric} & \textbf{Our Model} & \textbf{Baseline} \\
    \midrule
    \multirow{4}{*}{MDD}  
    & Accuracy   & $0.71$ & $0.63$ \\
    & Recall     & $0.81$ & $0.60$ \\
    & Precision  & $0.75$ & $0.65$ \\
    & F1         & $0.78$ & $0.62$ \\
    \hline
    \multirow{4}{*}{AD}  
    & Accuracy   & $0.73$ & $0.64$ \\
    & Recall     & $0.70$ & $0.50$ \\
    & Precision  & $0.70$ & $0.32$ \\
    & F1         & $0.70$ & $0.39$ \\
    \hline
    \multirow{4}{*}{PD}  
    & Accuracy   & $0.49$ & $0.48$ \\
    & Recall     & $0.90$ & $0.49$ \\
    & Precision  & $0.48$ & $0.24$ \\
    & F1         & $0.63$ & $0.32$ \\
    \bottomrule
  \end{tabular}
\end{table}

We perform experiments on all three datasets.
Table~\ref{tab:result} shows the results. 
To demonstrate the performance of our method, we use a convolutional recurrent neural network (CRNN) model as a baseline for comparison.
This model consists of several convolution blocks, including convolution layers, batch normalization and max pooling, followed by a two-layer bidirectional long short-term memory for temporal modeling, and a final fully connected layer for classification.
The CRNN architecture is widely used for speech disease detection.
Experimental results reveal that our model demonstrates strong performance across multiple tasks, achieving high accuracy in detecting various diseases.
With its prefix-tuning approach, it effectively leverages the pre-trained generative speech language model, allowing it to utilize existing knowledge without requiring modifications to the model architecture or the training of new models.
This approach is particularly advantageous when working with limited data, as it enables the model to generalize well across different diseases without the need for large datasets.
Additionally, the ability to build on a pre-trained model proves beneficial in scenarios with diverse or extensive datasets, where the advantages of a larger, more generalized model may outweigh the specificity of a disease-specific model like CRNN.
Further tuning or adaptation could enhance the model's performance, aligning its flexibility and ability to handle data-scarce environments with improved disease-specific results.

It is worth noting that there is a relatively higher proportion of false positives in the PD dataset.
This can be attributed to the fact that PD consists entirely of speech recordings from Parkinson's disease patients, where the label assignment considers unmedicated instances as positive and medicated instances as negative.
Even under medication, patients with Parkinson's disease can still exhibit certain symptoms, leading to the presence of pathological features in recorded speeches.

This observation highlights the challenges associated with accurately classifying PD speech samples due to the potential persistence of pathological characteristics, despite medication.
The higher false positive rate in the PD dataset can be attributed to the inherent complexities of the disease and the presence of residual symptoms in medicated speech recordings.
Nevertheless, the overall performance of our framework demonstrates its robustness and suitability for the task at hand.
Further improvements and refinements can be explored to address the specific challenges posed by the PD dataset.


\begin{table}[h]
\centering
\caption{Results before and after voting. }
\label{tab:voting}
\begin{tabular}{c c c}
\toprule
\textbf{Dataset} & \textbf{w/o Voting} & \textbf{Voting} \\
\midrule
MDD & 0.60 & 0.71 \\
AD & 0.58 & 0.73 \\
PD & 0.46 & 0.49 \\
\bottomrule
\end{tabular}
\end{table}

We also compare classification results before and after the voting process.
It can be seen from Table~\ref{tab:voting} that the voting technique greatly improves accuracy for MDD, AD and PD datasets.
This indicates that within the long speech recordings of each participant, different segments contribute unequally to the detection of the disorder.
By using the voting method, we effectively overcome this imbalance, resulting in a significant improvement in the final detection performance.


\section{Conclusions}
In this study, we have tackled the challenge of analyzing pathological speech unifiedly by employing prompt-tuning techniques to efficiently fine-tune a language model, validating its effectiveness through experiments on three diverse datasets. 
In our prompt tuning method, we use a well-crafted prefix prompt to direct the unit language model's inference. It consists of three main elements: disease-specific, language, and class embeddings. 
This framework utilized specialized spoken language models specifically designed for pathological speech, allowing for a greater focus on the unique features of pathological speech itself.
Another key contribution of our approach is the integration of a voting mechanism, which significantly improved classification results in diagnostic interviews involving various pathologies. 
With the utilization of improved language models and the adaptation to a wider range of pathological speech characteristics, unified pathological analysis system can better share knowledge between different diseases, leading to a more profound utilization of medical data and a more efficient cross-data training process.


\bibliographystyle{IEEEtran}
\bibliography{mybib}

\begin{thebibliography}{10}
\providecommand{\url}[1]{#1}
\csname url@samestyle\endcsname
\providecommand{\newblock}{\relax}
\providecommand{\bibinfo}[2]{#2}
\providecommand{\BIBentrySTDinterwordspacing}{\spaceskip=0pt\relax}
\providecommand{\BIBentryALTinterwordstretchfactor}{4}
\providecommand{\BIBentryALTinterwordspacing}{\spaceskip=\fontdimen2\font plus
\BIBentryALTinterwordstretchfactor\fontdimen3\font minus
  \fontdimen4\font\relax}
\providecommand{\BIBforeignlanguage}[2]{{%
\expandafter\ifx\csname l@#1\endcsname\relax
\typeout{** WARNING: IEEEtran.bst: No hyphenation pattern has been}%
\typeout{** loaded for the language `#1'. Using the pattern for}%
\typeout{** the default language instead.}%
\else
\language=\csname l@#1\endcsname
\fi
#2}}
\providecommand{\BIBdecl}{\relax}
\BIBdecl

\bibitem{CUMMINS201841}
\BIBentryALTinterwordspacing
N.~Cummins, A.~Baird, and B.~W. Schuller, ``Speech analysis for health: Current
  state-of-the-art and the increasing impact of deep learning,''
  \emph{Methods}, vol. 151, pp. 41--54, 2018, health Informatics and
  Translational Data Analytics. [Online]. Available:
  \url{https://www.sciencedirect.com/science/article/pii/S1046202317303717}
\BIBentrySTDinterwordspacing

\bibitem{10.1145/3560815}
\BIBentryALTinterwordspacing
P.~Liu, W.~Yuan, J.~Fu, Z.~Jiang, H.~Hayashi, and G.~Neubig, ``Pre-train,
  prompt, and predict: A systematic survey of prompting methods in natural
  language processing,'' \emph{ACM Comput. Surv.}, vol.~55, no.~9, jan 2023.
  [Online]. Available: \url{https://doi.org/10.1145/3560815}
\BIBentrySTDinterwordspacing

\bibitem{li-liang-2021-prefix}
\BIBentryALTinterwordspacing
X.~L. Li and P.~Liang, ``Prefix-tuning: Optimizing continuous prompts for
  generation,'' in \emph{Proceedings of the 59th Annual Meeting of the
  Association for Computational Linguistics and the 11th International Joint
  Conference on Natural Language Processing (Volume 1: Long Papers)}, C.~Zong,
  F.~Xia, W.~Li, and R.~Navigli, Eds.\hskip 1em plus 0.5em minus 0.4em\relax
  Online: Association for Computational Linguistics, Aug. 2021, pp. 4582--4597.
  [Online]. Available: \url{https://aclanthology.org/2021.acl-long.353}
\BIBentrySTDinterwordspacing

\bibitem{chang2023speechprompt}
K.-W. Chang, Y.-K. Wang, H.~Shen, I.~thing Kang, W.-C. Tseng, S.-W. Li, and
  H.~yi~Lee, ``Speechprompt v2: Prompt tuning for speech classification
  tasks,'' 2023.

\bibitem{chang2022speechprompt}
K.-W. Chang, W.-C. Tseng, S.-W. Li, and H.-y. Lee, ``Speechprompt: An
  exploration of prompt tuning on generative spoken language model for speech
  processing tasks,'' \emph{arXiv preprint arXiv:2203.16773}, 2022.

\bibitem{hsu2021hubert}
W.-N. Hsu, B.~Bolte, Y.-H.~H. Tsai, K.~Lakhotia, R.~Salakhutdinov, and
  A.~Mohamed, ``Hubert: Self-supervised speech representation learning by
  masked prediction of hidden units,'' \emph{IEEE/ACM Transactions on Audio,
  Speech, and Language Processing}, vol.~29, pp. 3451--3460, 2021.

\bibitem{lakhotia2021generative}
K.~Lakhotia, E.~Kharitonov, W.-N. Hsu, Y.~Adi, A.~Polyak, B.~Bolte, T.-A.
  Nguyen, J.~Copet, A.~Baevski, A.~Mohamed \emph{et~al.}, ``On generative
  spoken language modeling from raw audio,'' \emph{Transactions of the
  Association for Computational Linguistics}, vol.~9, pp. 1336--1354, 2021.

\bibitem{ott2019fairseq}
M.~Ott, S.~Edunov, A.~Baevski, A.~Fan, S.~Gross, N.~Ng, D.~Grangier, and
  M.~Auli, ``fairseq: A fast, extensible toolkit for sequence modeling,'' in
  \emph{Proceedings of NAACL-HLT 2019: Demonstrations}, 2019.

\bibitem{DI2021161}
\BIBentryALTinterwordspacing
Y.~Di, J.~Wang, W.~Li, and T.~Zhu, ``Using i-vectors from voice features to
  identify major depressive disorder,'' \emph{Journal of Affective Disorders},
  vol. 288, pp. 161--166, 2021. [Online]. Available:
  \url{https://www.sciencedirect.com/science/article/pii/S0165032721003281}
\BIBentrySTDinterwordspacing

\bibitem{li2019building}
X.~Li, Q.~Qiu, Y.~Yang, L.~Sun, M.~Jiang, c.~Gu, M.~Cui, and X.~Lin, ``Building
  a continuous dementia management model in communities of shanghai,''
  \emph{Innovation in Aging}, vol.~3, no. Supplement\_1, pp. S444--S444, 2019.

\bibitem{Zhang2023ReCLRRC}
\BIBentryALTinterwordspacing
P.~Zhang, M.~Wu, and K.~Yu, ``Reclr: Reference-enhanced contrastive learning of
  audio representation for depression detection,'' \emph{INTERSPEECH 2023},
  2023. [Online]. Available:
  \url{https://api.semanticscholar.org/CorpusID:260910099}
\BIBentrySTDinterwordspacing

\bibitem{PrezToro2023AutomaticAO}
\BIBentryALTinterwordspacing
P.~A. P{\'e}rez-Toro, T.~Arias-Vergara, F.~Braun, F.~H{\"o}nig, C.~A.
  T{\'o}bon-Quintero, D.~Aguill{\'o}n, F.~Lopera, L.~Hincapi{\'e}-Henao,
  M.~Schuster, K.~Riedhammer, A.~K. Maier, E.~N{\"o}th, and J.~R.
  Orozco-Arroyave, ``Automatic assessment of alzheimer's across three languages
  using speech and language features,'' \emph{INTERSPEECH 2023}, 2023.
  [Online]. Available: \url{https://api.semanticscholar.org/CorpusID:260919539}
\BIBentrySTDinterwordspacing

\bibitem{gratch2014distress}
J.~Gratch, R.~Artstein, G.~M. Lucas, G.~Stratou, S.~Scherer, A.~Nazarian,
  R.~Wood, J.~Boberg, D.~DeVault, S.~Marsella, and D.~R. Traum, ``The distress
  analysis interview corpus of human and computer interviews,'' in
  \emph{Proceedings of the 9th International Conference on Language Resources
  and Evaluation (LREC)}, 2014.

\bibitem{vazquez2020automatic}
A.~V{\'a}zquez-Romero and A.~Gallardo-Antol{\'\i}n, ``Automatic detection of
  depression in speech using ensemble convolutional neural networks,''
  \emph{Entropy}, vol.~22, no.~6, p. 688, 2020.

\bibitem{valstar2016summary}
M.~Valstar, J.~Gratch, B.~Schuller, F.~Ringeval, R.~Cowie, and M.~Pantic,
  ``Summary for avec 2016: Depression, mood, and emotion recognition workshop
  and challenge,'' in \emph{Proceedings of the 24th ACM international
  conference on Multimedia}, 2016, pp. 1483--1484.

\bibitem{REJAIBI2022103107}
\BIBentryALTinterwordspacing
E.~Rejaibi, A.~Komaty, F.~Meriaudeau, S.~Agrebi, and A.~Othmani, ``Mfcc-based
  recurrent neural network for automatic clinical depression recognition and
  assessment from speech,'' \emph{Biomedical Signal Processing and Control},
  vol.~71, p. 103107, 2022. [Online]. Available:
  \url{https://www.sciencedirect.com/science/article/pii/S1746809421007047}
\BIBentrySTDinterwordspacing

\bibitem{ZOLNOORI2023102624}
\BIBentryALTinterwordspacing
M.~Zolnoori, A.~Zolnour, and M.~Topaz, ``Adscreen: A speech processing-based
  screening system for automatic identification of patients with alzheimer's
  disease and related dementia,'' \emph{Artificial Intelligence in Medicine},
  vol. 143, p. 102624, 2023. [Online]. Available:
  \url{https://www.sciencedirect.com/science/article/pii/S0933365723001380}
\BIBentrySTDinterwordspacing

\bibitem{liu2023efficient}
J.~Liu, F.~Fu, L.~Li, J.~Yu, D.~Zhong, S.~Zhu, Y.~Zhou, B.~Liu, and J.~Li,
  ``Efficient pause extraction and encode strategy for alzheimer’s disease
  detection using only acoustic features from spontaneous speech,'' \emph{Brain
  Sciences}, vol.~13, no.~3, p. 477, 2023.

\bibitem{liu2023ensemble}
Z.~Liu, H.~Yu, G.~Li, Q.~Chen, Z.~Ding, L.~Feng, Z.~Yao, and B.~Hu, ``Ensemble
  learning with speaker embeddings in multiple speech task stimuli for
  depression detection,'' \emph{Frontiers in Neuroscience}, vol.~17, p.
  1141621, 2023.

\bibitem{lamba2023speech}
R.~Lamba, T.~Gulati, A.~Jain, and P.~Rani, ``A speech-based hybrid decision
  support system for early detection of parkinson's disease,'' \emph{Arabian
  Journal for Science and Engineering}, vol.~48, no.~2, pp. 2247--2260, 2023.

\bibitem{bertini2022automatic}
F.~Bertini, D.~Allevi, G.~Lutero, L.~Calz{\`a}, and D.~Montesi, ``An automatic
  alzheimer’s disease classifier based on spontaneous spoken english,''
  \emph{Computer Speech \& Language}, vol.~72, p. 101298, 2022.

\bibitem{quan2022end}
C.~Quan, K.~Ren, Z.~Luo, Z.~Chen, and Y.~Ling, ``End-to-end deep learning
  approach for parkinson’s disease detection from speech signals,''
  \emph{Biocybernetics and Biomedical Engineering}, vol.~42, no.~2, pp.
  556--574, 2022.

\bibitem{yin2023depression}
F.~Yin, J.~Du, X.~Xu, and L.~Zhao, ``Depression detection in speech using
  transformer and parallel convolutional neural networks,'' \emph{Electronics},
  vol.~12, no.~2, p. 328, 2023.

\bibitem{jahan2024early}
Z.~Jahan, S.~B. Khan, and M.~Saraee, ``Early dementia detection with speech
  analysis and machine learning techniques,'' \emph{Discover Sustainability},
  vol.~5, no.~1, pp. 1--18, 2024.

\bibitem{koops2023speech}
S.~Koops, S.~G. Brederoo, J.~N. de~Boer, F.~G. Nadema, A.~E. Voppel, and I.~E.
  Sommer, ``Speech as a biomarker for depression,'' \emph{CNS \& Neurological
  Disorders-Drug Targets (Formerly Current Drug Targets-CNS \& Neurological
  Disorders)}, vol.~22, no.~2, pp. 152--160, 2023.

\end{thebibliography}

\end{document}